\documentclass[%
reprint,
showpacs,preprintnumbers,
 amsmath,amssymb,
aip,
]{revtex4-1}
\usepackage{graphicx}
\usepackage{dcolumn}
\usepackage{bm}
\usepackage{color}

\begin{document}


\title{Stabilization of helical magnetic structures in thin multilayers}

\author{L.\,V.~Dzemiantsova}
 \email{liudmila.dzemiantsova@desy.de}
  \affiliation{%
 The Hamburg Centre for Ultrafast Imaging, Luruper Chaussee 149, 22761 Hamburg, Germany
}%
   \affiliation{%
 Deutsches Elektronen-Synchrotron, Notkestra$\beta$e 85, 22607 Hamburg, Germany
}%
\author{G.~Meier}
\affiliation{%
 The Hamburg Centre for Ultrafast Imaging, Luruper Chaussee 149, 22761 Hamburg, Germany
}
\affiliation{%
 Max-Planck Institute for the Structure and Dynamics of Matter, Luruper Chaussee 149, 22761 Hamburg, Germany
}%
\author{R.~R\"{o}hlsberger}
\affiliation{%
 The Hamburg Centre for Ultrafast Imaging, Luruper Chaussee 149, 22761 Hamburg, Germany
}%
\affiliation{%
 Deutsches Elektronen-Synchrotron, Notkestra$\beta$e 85, 22607 Hamburg, Germany
}%

\date{\today}

\begin{abstract}
Based on micromagnetic simulations, we report on a novel helical magnetic structure in a soft magnetic film that is sandwiched between and exchange-coupled to two hard magnetic layers. Confined between antiparallel hard magnetic moments, a helix with a turn of \mbox{180$^{\circ}$} is stable without the presence of an external magnetic field. The magnetic stability is determined by the energy minimization and is a result of an internal field created by exchange interaction and anisotropy. Since the internal field stores magnetic energy, the helix can serve as an energy-storing element in spin-based nanodevices. Due to the significantly different magnetic resonance frequencies, the ferromagnetic and helical ground states are easy to distinguish in a broadband ferromagnetic resonance experiment.
\end{abstract}

\maketitle

Robust and energetically efficient magnetic structures that employ the spin degree of freedom to store and process information are at the heart of modern spin-based technology. Many experiments have been performed to investigate the interaction of spins with charges or external fields, using different device geometries like mechanically or lithographically fabricated point contacts, nanopillars or tunnel junctions.\cite{kuhlmann_2013,lehndorff_2008,krause_2007,zabel_2013} It has recently been shown that the transmission and processing of information without electric currents or external fields can be achieved via the spin degree of freedom subjected to exchange, Ruderman-Kittel-Kasuya-Yosida (RKKY) or long-range dipolar interactions.\cite{vedmedenko_2014} When structural boundaries fix the magnetization, these interactions can topologically stabilize configurations like spin helices. Though a large variety of devices with desirable parameters can be artificially fabricated down to the sub-nanoscale, the creation of helices with stable magnetic properties can, however, be an experimental challenge. 

Here, we propose an approach for creating novel helical magnetic structures in thin multilayer systems. We show that such structures initially twisted in an external magnetic field, stay stable even without the presence of the field. In contrast to rare earth materials where a helical order is governed by the RKKY interaction,\cite{jensen_1991,blundell_2009} the functionality of the system in this study relies critically on the exchange-coupling mechanism of thin layers consisting of a hard and a soft-magnetic material. As a characteristic property of exchange-coupled layers or exchange spring magnets, the magnetization of the soft-magnetic film is pinned to the hard-magnetic film at the interface as a result of the exchange interaction.\cite{rohlsberger_2002} With increasing distance from the interface, the exchange coupling becomes weaker and the magnetic moments in the soft layer form a spiral structure under the action of an external field. To stabilize this spin spiral structure, we add on top a magnetic film with an anisotropy that lies in between those of the hard and soft materials. When the external field is removed, such a trilayer can relax in a new stable configuration where a helical magnetic order exists. The magnetic stability is the result of an internal effective field which is created by the exchange interaction and the anisotropy and stores magnetic energy. This field causes ferromagnetic resonance (FMR) frequencies of the magnetization precession in the GHz range, which are higher compared to those when the films show a ferromagnetic order at zero applied magnetic fields.

Nanocomposite materials with a stable helical order open broad perspectives for future scientific and technological applications in the field of spin engineering on smallest length scales. Since these structures store magnetic energy, they can serve as energy-storing elements in modern spin-based nanodevices. The generation of high-frequency signals without a presence of external fields or currents renders magnetic helices promising candidates for application in field- or charge-free spin-transfer effects in nanoscale magnetic schemes. 

\begin{figure}[t]
\centering
\includegraphics[width=1.0\linewidth]{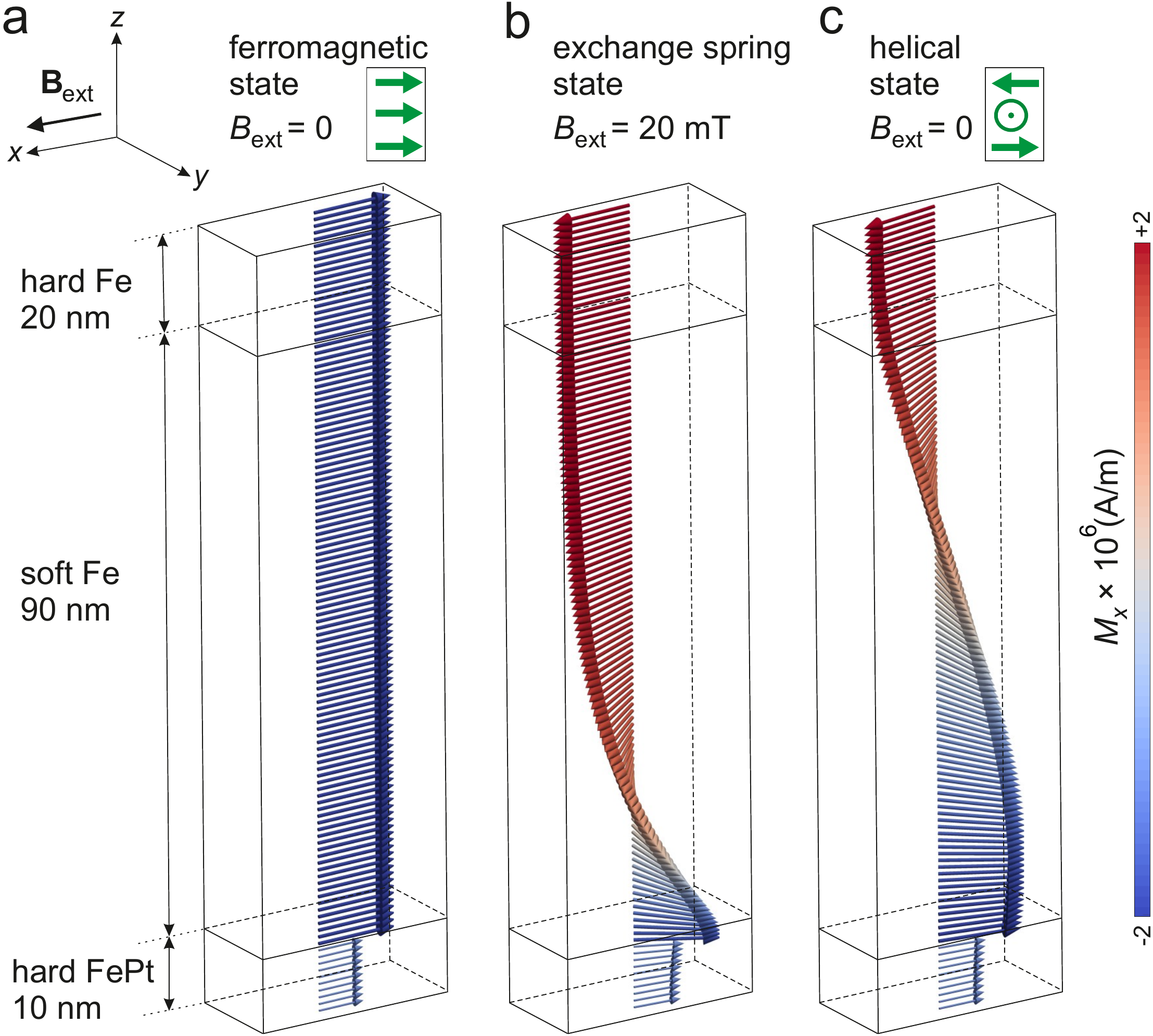}
\caption{\label{fig:fig_1}{(color online) 3D representation of micromagnetic simulation data for a \mbox{hard FePt/soft Fe/hard Fe} trilayer with localized magnetic moments (arrows inside slabs) ordered in the (a) ferromagnetic, (b) exchange spring and (c) helical state. As the initial ground state, the trilayer shows the ferromagnetic order at ${B}_\textnormal{ext}=0$. To create the exchange spring, an external magnetic field ${B}_\textnormal{ext}\textgreater0$ has to be applied. The helical state is the second ground state that can be achieved when the field is removed. The colorscale highlights the \mbox{$x$-component} of the magnetic moments, $M_x$, with respect to minimum (blue) and maximum (red) values. }}
\end{figure}

Our approach is described within a one-dimensional micromagnetic model utilizing the MicroMagnum code that computes the Landau-Lifshitz-Gilbert (LLG) equation.\cite{magnum} The effective magnetic energy ${E}_\textnormal{eff}$ in the LLG equation is the sum of the exchange ${E}_\textnormal{exch}$, demagnetization ${E}_\textnormal{dem}$, anisotropy ${E}_\textnormal{anis}$ and external magnetic ${E}_\textnormal{ext}$ energies. In the present study, we consider a trilayer with lateral dimensions of $10\times10$~$\mu$$\textnormal{m}^{2}$ size with fixed thickness. The film is discretized into 1-nm thick platelets, and each platelet is assumed to have an uniaxial easy magnetization axis along the in-plane $x$-direction (see \mbox{Fig.~\,\ref{fig:fig_1}(a)}). To realize a helical magnetic order, we model a layered system, where the individual layers differ in thickness $d$, magnetization saturation $M_\textnormal{s}$ and magnetic anisotropy $K$. As shown in \mbox{Fig.~\,\ref{fig:fig_1}(a)}, the system consists of three magnets:\\\mbox{($i$) a hard magnetic FePt film (\mbox{$d_\textnormal{1}=10$ nm},} \mbox{$M_\textnormal{s1}=11.0\times10^{5}$ A/m}, \mbox{$K_\textnormal{1}=4.4\times10^{5}$~J/$\textnormal{m}^{3}$}),\cite{anisotropy}\\\mbox{($ii$) a soft magnetic Fe film (\mbox{$d_\textnormal{2}=90$~nm},} \mbox{$M_\textnormal{s2}=19.0\times10^{5}$~A/m}, \mbox{$K_\textnormal{2}=1.0\times10^{2}$~J/$\textnormal{m}^{3}$}) and\\\mbox{($iii$) a hard magnetic Fe film (\mbox{$d_\textnormal{3}=20$~nm},} \mbox{$M_\textnormal{s3}=19.0\times10^{5}$ A/m}, \mbox{$K_\textnormal{3}=7.0\times10^{4}$~J/$\textnormal{m}^{3}$}).\cite{experiment}\\The exchange constants of all materials are chosen to be \mbox{$A=1.0\times10^{-11}$~J/m}, according to Ref.~\onlinecite{rohlsberger_2002}. The Gilbert damping constants for FePt and both Fe films are 0.02 and 0.01, respectively. As an initial ground state, the composite shows a ferromagnetic order (Fig.~\,\ref{fig:fig_1}(a)).

\begin{figure}[t]
\centering
\includegraphics[width=1\linewidth]{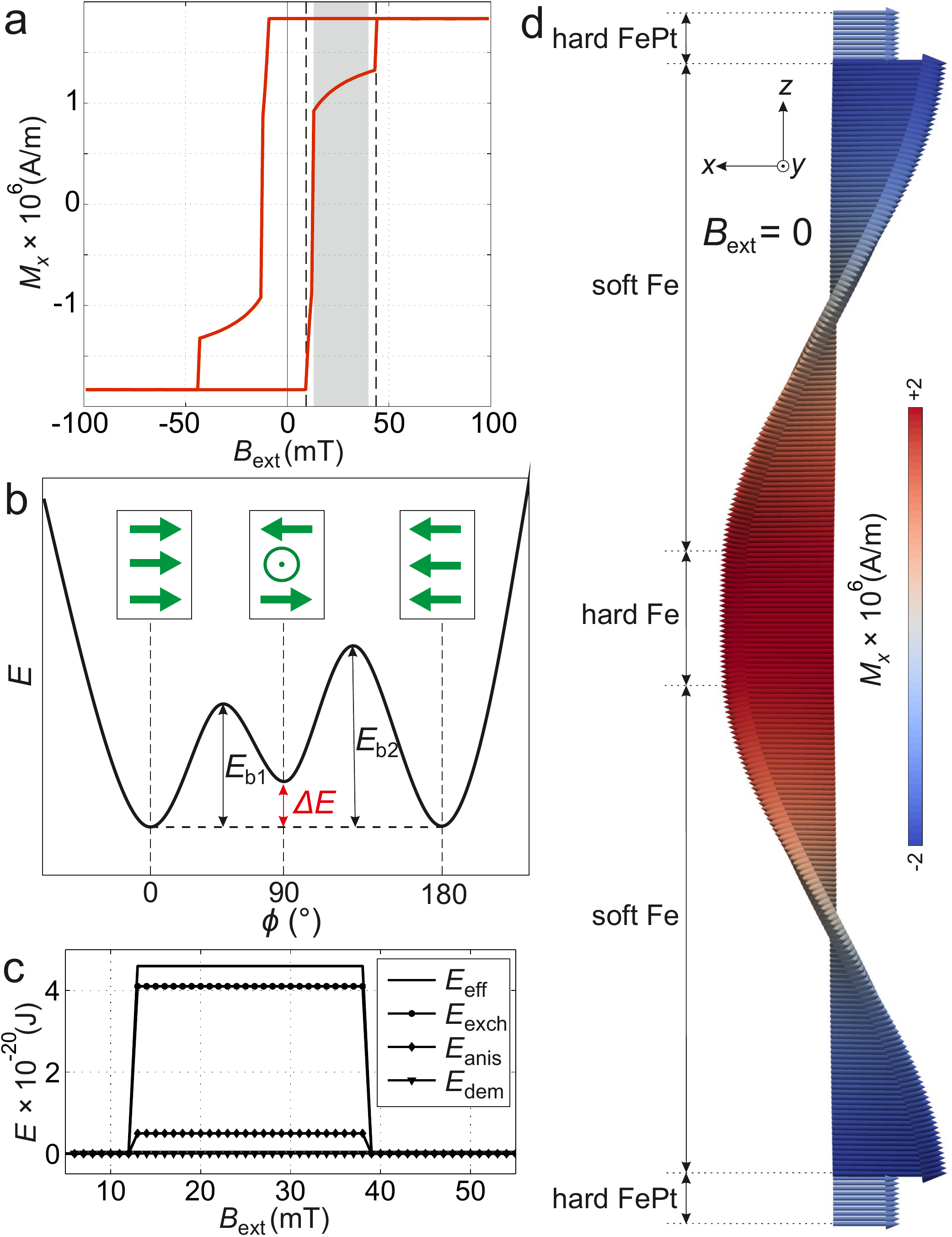} 
\caption{\label{fig:fig_2} {(color online) (a) Hysteresis loop of the trilayer. Dashed lines indicate a region of applied fields where the structure is in the exchange spring state shown in Fig.~\,\ref{fig:fig_1}(b). The gray area maps out a field range where the system is in the helical state shown in Fig.~\,\ref{fig:fig_1}(c) after the external field is removed. (b) Schematic drawing of energy barriers ${E}_{\textnormal{b}i}$ ($i$~=~1,~2) at ${B}_\textnormal{ext}=0$. Two global energy minima at \mbox{$0^{\circ}$} and \mbox{$180^{\circ}$} correspond to a ferromagnetic state, a local minimum at \mbox{$90^{\circ}$} corresponds to a helical state. $\phi$ is the angle between the mid magnetic moment in the soft Fe film and the -$x$-axis. $\Delta{E}$ is the energy stored in the helix. (c) Contributions of ${E}_\textnormal{exch}$, ${E}_\textnormal{anis}$ and ${E}_\textnormal{dem}$ to ${E}_\textnormal{eff}$, when the trilayer is in an energy equilibrium at ${B}_\textnormal{ext}=0$. Energies are plotted as functions of ${B}_\textnormal{ext}$ that was applied before relaxation to magnetize films. (d) A double-helix constructed from two single helices that are placed up-side-down on top of each other. The colorscale highlights the $x$-component of the magnetic moments, $M_x$, with respect to minimum (blue) and maximum (red) values.}}
\end{figure}

To find an equilibrium state other than the initial state, we repeat the following procedure for increasing values of an external magnetic field ${B}_\textnormal{ext}$: Starting with \mbox{1~mT}, we apply ${B}_\textnormal{ext}$ opposite to the initial direction of the magnetic moments in FePt and relax the trilayer. The field is then set to zero and the structure is again relaxed. The procedure is repeated for \mbox{2~mT}, \mbox{3~mT} up to \mbox{100~mT}. At each step, we observe the magnetic state after the relaxation in the external field and subsequently without it. Under the applied external field, the magnetic moments in the Fe films are spatially twisted due to exchange-coupling between FePt and Fe through the interface (see Fig.~\,\ref{fig:fig_1}(b)). We find that the exchange spring configuration exists at fields of 9~up to~44~mT as indicated by the dashed lines in the hysteresis loop in Fig.~\,\ref{fig:fig_2}(a). With further increase of the field strength the magnetic moments of the system completely switch by \mbox{180$^{\circ}$}. 

Depending on the energy ${E}_\textnormal{ext}$ introduced by the external field, the layer stack can relax into ground states with different energy minima when the field is removed. As schematically illustrated in Fig.~\,\ref{fig:fig_2}(b), energy barriers ${E}_{\textnormal{b}i}$ ($i$~=~1,~2) separate two global minima ${E}_\textnormal{min,0}$ at \mbox{$\phi=0^{\circ}$} and ${E}_\textnormal{min,180}$ at \mbox{$\phi=180^{\circ}$} and one local minimum ${E}_\textnormal{min,90}$ at \mbox{$\phi=90^{\circ}$} with $\phi$ being the angle between the mid magnetic moment in the soft Fe film and the \mbox{-$x$-axis}. If \mbox{${E}_\textnormal{ext}$ $\textless$ ${E}_{\textnormal{b}1}$} or \mbox{${E}_\textnormal{ext}$ $\textgreater$ ${E}_{\textnormal{b}2}$}, the system relaxes into a stable state with the global energy minimum at \mbox{$\phi=0^{\circ}$} or \mbox{$\phi=180^{\circ}$}, respectively. In these states the magnetic moments of the films are aligned parallel (\mbox{$\phi=0^{\circ}$}) or antiparallel (\mbox{$\phi=180^{\circ}$}) to those of the initial state. In case of \mbox{${E}_{\textnormal{b}1}$ $\textless$ ${E}_\textnormal{ext}$ $\textless$ ${E}_{\textnormal{b}2}$}, the system relaxes into the metastable state with the local energy minimum at \mbox{$\phi=90^{\circ}$}. We find that this state corresponds to a magnetic configuration where the magnetic moments in the soft Fe film show a helical order with a turn of \mbox{$180^{\circ}$} and are pinned between the antiparallel magnetizations of the hard FePt film and the hard Fe film at the ends (see Fig.~\,\ref{fig:fig_1}(c)). We observe the relaxed helical configuration after the magnetic moments were twisted in external fields in the range of 13~to~38~mT as mapped out by the gray area in Fig.~\,\ref{fig:fig_2}(b). 

The reason for the stability of the magnetic helices at ${B}_\textnormal{ext}=0$ is an internal effective magnetic field in the soft layer. To create the internal field, following conditions have to be fulfilled: ($i$)~an external field that introduces energy into the system; ($ii$) and ($iii$)~magnetic moments at the ends of the structure are aligned antiparallel and are fixed, respectively. Under the action of the external field, the magnetization of the hard Fe film can be twisted by \mbox{$180^{\circ}$} whereas the magnetization of the hard FePt film stays unchanged due to the large anisotropy. When the external field is removed, the magnetization of the hard Fe film does not reverse since its anisotropy is large compared to the exchange interaction that tends to unwind the exchange spring and is weaker with distance from the hard FePt magnet underneath. Confined between antiparallel magnetizations, the magnetic moments in the soft Fe film relax into a stable configuration where they continuously change their direction from parallel to antiparallel, i.e. exhibiting a helical order.\cite{vedmedenko_2014} At the energy equilibrium the internal field is created that stores exchange and anisotropy energies via \mbox{${E}_\textnormal{eff}={E}_\textnormal{exch}+{E}_\textnormal{anis}$}, and is equal to the energy difference \mbox{$\Delta{E}={E}_{\textnormal{min},i}-{E}_\textnormal{min,90}$ ($i$~=~0,~180)} (see Fig.~\,\ref{fig:fig_2}(b)). With ${E}_\textnormal{eff}$, ${E}_\textnormal{exch}$, ${E}_\textnormal{anis}$ and ${E}_\textnormal{dem}$ plotted as functions of ${B}_\textnormal{ext}$ that was applied to magnetize the films (Fig.~\,\ref{fig:fig_2}(c)), we find that ${E}_\textnormal{exch}$ makes the main contribution to ${E}_\textnormal{eff}$. This means that the exchange interaction is mostly responsible for the creation of the internal effective magnetic field. 

The multilayers can store energy \mbox{$\Delta{E}=4.6\times10^{-20}$~J} that is about eleven times more than the thermal activation energy \mbox{$k_\textnormal{B}T=4.1\times10^{-21}$~J} with \mbox{$k_\textnormal{B}=13.8\times10^{-24}$~J/K} being the Boltzmann constant and \mbox{$T=300$~K} being the temperature. The same value of $\Delta{E}$ can be obtained when one employs soft FePt (\mbox{$d_\textnormal{1}=30$~nm}, \mbox{$M_\textnormal{s1}=11.0\times10^{5}$~A/m}, \mbox{$K_\textnormal{1}=4.4\times10^{4}$~J/$\textnormal{m}^{3}$}) instead of the hard Fe film, or \mbox{permalloy \mbox{Ni$_\textnormal{80}$Fe$_\textnormal{20}$} (\mbox{$d_\textnormal{2}=90$~nm}, \mbox{$M_\textnormal{s2}=8.6\times10^{5}$~A/m},} \mbox{$K_\textnormal{2}=1.0\times10^{2}$~J/$\textnormal{m}^{3}$}, \mbox{$A_\textnormal{2}=1.3\times10^{-11}$~J/m}) which can replace the soft Fe magnet. The stored energy can be increased by a factor of two in case of a double-helix where the trilayers are duplicated and placed up-side-down on top of each other as show in Fig.~\,\ref{fig:fig_2}(d). A significant amount of stored energy can be reached when the magnetic hardness of the underlying and top magnets is increased. For example, if the anisotropy constants of the hard films \mbox{$K_\textnormal{1}$} and \mbox{$K_\textnormal{3}$} would be ten times larger,\cite{fullerton_1997} the amount of stored energy could be increased by a factor of four after magnetizing the films in ${B}_\textnormal{ext}=200$~mT. The thickness of the soft Fe film in such a composite must be a third of that shown in Fig.~\,\ref{fig:fig_1} to obtain the helical configuration.

\begin{figure}[t]
\centering
\includegraphics[width=1\linewidth]{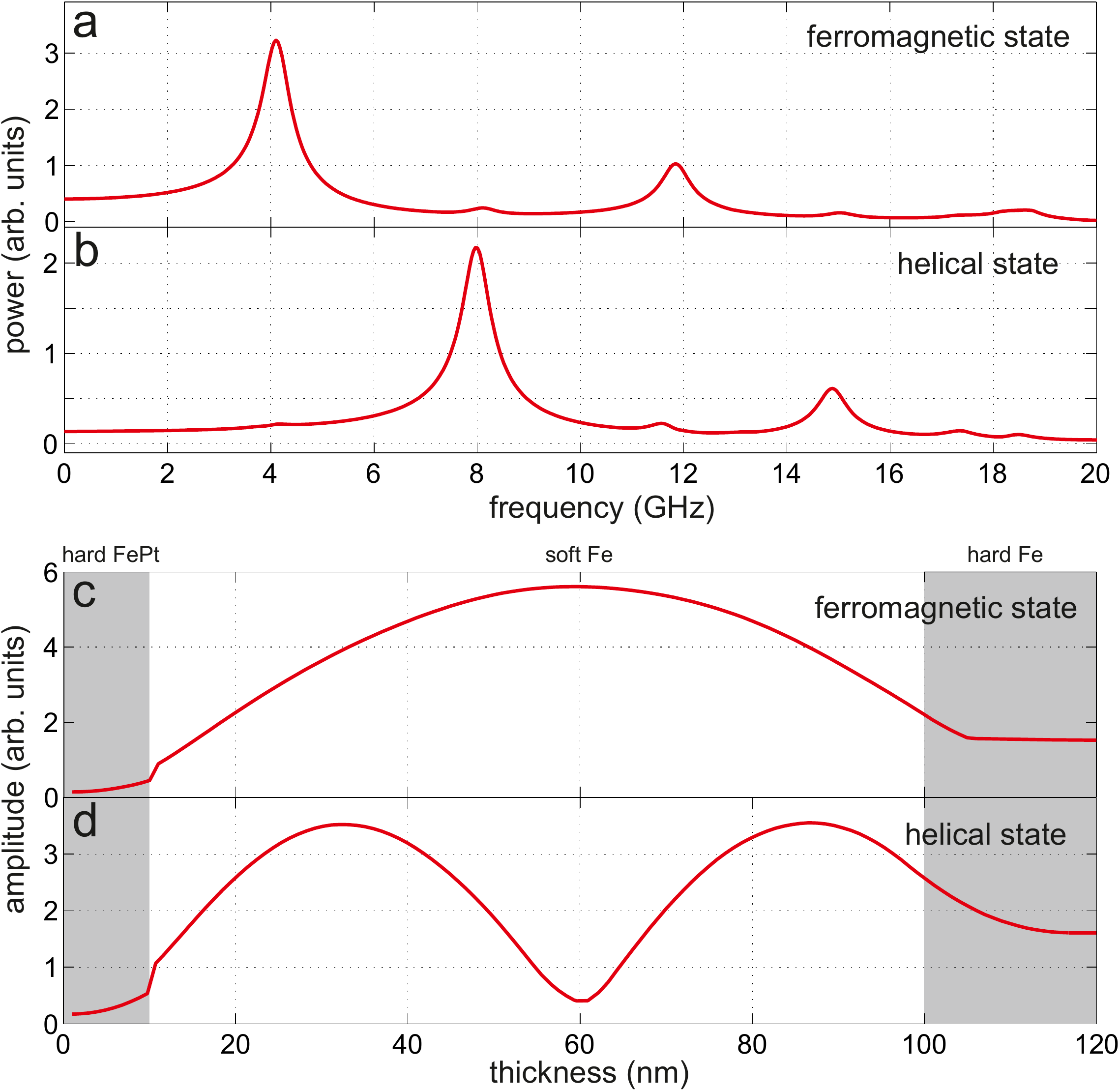}
\caption{\label{fig:fig_3}{(color online) FMR spectra of the trilayer in the (a) ferromagnetic and (b) helical state. Due to the internal magnetic field, resonance frequencies in the helix are higher compared to those in the ferromagnetic case. A spatial profile of the lowest frequency excitation is shown in (c) for the ferromagnetic and in (d) for the helical state. In all cases, ${B}_\textnormal{ext}=0$}.}
\end{figure}

The internal field can have an effect on the precessional motion of the magnetization, which is crucial for transmitting and processing information in thin multilayers via spin waves. We explore the magnetization dynamics in the trilayer in the ferromagnetic (FM) and the helical magnetic (HM) state after a magnetic pulse of 0.4~ns duration and 1~mT amplitude is applied in $z$-direction. This field direction yields the strongest response. The precessional motion is simulated for 3~ns at ${B}_\textnormal{ext}=0$. To obtain a frequency spectrum of the oscillations, we analyze the dynamic response by a Fast Fourier transformation (FFT) of the $z$-component of the magnetization, averaged over all layers. As shown in \mbox{Fig.~\,\ref{fig:fig_3}(a) and (b)}, the FMR spectra of the trilayer show two main resonant peaks in the range of 0 to 20~GHz. Since the first (lowest) frequency resonance is closest to a spatially uniform mode, it shows the largest amplitude.\cite{stancil_2009} We find that in case of the HM state, FMR frequencies (\mbox{${f}_\textnormal{HM,1}=8.0$~GHz} and \mbox{${f}_\textnormal{HM,2}=14.9$~GHz}) are higher compared to a film with the FM order (\mbox{${f}_\textnormal{FM,1}=4.1$~GHz} and \mbox{${f}_\textnormal{FM,2}=11.8$~GHz}). In particular, ${f}_\textnormal{HM,1}$ is about a factor of two higher than ${f}_\textnormal{FM,1}$. It is known that the frequency of magnetization precession depends on the strength of the applied magnetic field: as the external field strength is increased, the precession frequency increases.\cite{stancil_2009} In the helix, the internal field replaces the external field, therefore the increase of FMR frequencies compared to those in the ferromagnetic case at ${B}_\textnormal{ext}=0$ is expected. 

The internal field also modifies the spatial profiles of magnetic excitations. Figures~\,\ref{fig:fig_3}(c) and (d) show amplitudes of the lowest frequency resonance, plotted as functions of the thickness for the FM and the HM case, respectively. The excitations are known as perpendicular standing spin waves (PSSW).\cite{seki_2013} It can be seen in \mbox{Fig.~\,\ref{fig:fig_3}(c)} that the lowest frequency excitation is mainly localized in the Fe layers with a maximum at the thickness \mbox{$d=60$~nm}. In contrast to the FM case, the lowest frequency excitation in the helix has two maxima at \mbox{$d=30$~nm} and \mbox{$d=86$~nm} separated by a minimum at \mbox{$d=60$~nm}. This is a result of the twist of the mid magnetic moment perpendicular to the fixed layer magnetization, hence parallel to the hard axis of the trilayer. In order to excite magnetic moments along the hard axis, a stronger magnetic pulse is required. Therefore, the amplitude of excitations gradually reduces (increases) from maximum to minimum (from minimum to maximum) as magnetic moments rotate from \mbox{$0^{\circ}$} to \mbox{$90^{\circ}$} (from \mbox{$90^{\circ}$} to \mbox{$180^{\circ}$}). 

In conclusion, we show a pathway to create a helical spin order in thin hard FePt/soft Fe/hard Fe films, which is stable at zero applied external fields. The magnetic stability is determined by the energy minimization and depends on the thickness of the films, the relative strengths and directions of the anisotropies, as well as the exchange coupling between the magnetic layers. We observe the helical turn of \mbox{$180^{\circ}$} in the soft film, confined between the antiparallel magnetic moments of hard layers. Due to the internal field created mainly by exchange interaction, the helix stores magnetic energy. In case of the helix, FMR frequencies of the magnetization precession are higher compared to those when films show a ferromagnetic order at ${B}_\textnormal{ext}=0$. In particular, the ratio of the lowest resonance frequencies ${f}_\textnormal{HM,1}/{f}_\textnormal{FM,1}$ is approximately two. The spatial profile of the lowest frequency excitation shows two maxima separated by a minimum, which is in contrast to a single maximum in the middle of the trilayer in the ferromagnetic case. 

By measuring broadband FMR spectra with a vector network analyzer,\cite{swoboda_2013,kuhlmann_2013,zabel_2013} one can indirectly detect the helical order in an experiment. The direct observation can be performed via magnetic superstructure reflections that are widely used in resonant x-ray and polarized-neutron reflectometry to characterize the vertical spin profile in multilayer systems.\cite{schlage_2013} Structures with a stable helical configuration provide a new route to store, transmit and process information in the wide range of spintronic and recording media applications. 

We thank Lars Bocklage for fruitful discussions. Financial support from the Deutsche Forschungsgemeinschaft via excellence cluster ``The Hamburg Centre for Ultrafast Imaging - Structure, Dynamics and Control of Matter on the Atomic Scale'' is gratefully acknowledged.

\bibliography{helix}

\begin{thebibliography}{18}%
\makeatletter
\providecommand \@ifxundefined [1]{%
 \@ifx{#1\undefined}
}%
\providecommand \@ifnum [1]{%
 \ifnum #1\expandafter \@firstoftwo
 \else \expandafter \@secondoftwo
 \fi
}%
\providecommand \@ifx [1]{%
 \ifx #1\expandafter \@firstoftwo
 \else \expandafter \@secondoftwo
 \fi
}%
\providecommand \natexlab [1]{#1}%
\providecommand \enquote  [1]{``#1''}%
\providecommand \bibnamefont  [1]{#1}%
\providecommand \bibfnamefont [1]{#1}%
\providecommand \citenamefont [1]{#1}%
\providecommand \href@noop [0]{\@secondoftwo}%
\providecommand \href [0]{\begingroup \@sanitize@url \@href}%
\providecommand \@href[1]{\@@startlink{#1}\@@href}%
\providecommand \@@href[1]{\endgroup#1\@@endlink}%
\providecommand \@sanitize@url [0]{\catcode `\\12\catcode `\$12\catcode
  `\&12\catcode `\#12\catcode `\^12\catcode `\_12\catcode `\%12\relax}%
\providecommand \@@startlink[1]{}%
\providecommand \@@endlink[0]{}%
\providecommand \url  [0]{\begingroup\@sanitize@url \@url }%
\providecommand \@url [1]{\endgroup\@href {#1}{\urlprefix }}%
\providecommand \urlprefix  [0]{URL }%
\providecommand \Eprint [0]{\href }%
\providecommand \doibase [0]{http://dx.doi.org/}%
\providecommand \selectlanguage [0]{\@gobble}%
\providecommand \bibinfo  [0]{\@secondoftwo}%
\providecommand \bibfield  [0]{\@secondoftwo}%
\providecommand \translation [1]{[#1]}%
\providecommand \BibitemOpen [0]{}%
\providecommand \bibitemStop [0]{}%
\providecommand \bibitemNoStop [0]{.\EOS\space}%
\providecommand \EOS [0]{\spacefactor3000\relax}%
\providecommand \BibitemShut  [1]{\csname bibitem#1\endcsname}%
\let\auto@bib@innerbib\@empty
\bibitem [{\citenamefont {Kuhlmann}\ \emph {et~al.}(2013)\citenamefont
  {Kuhlmann}, \citenamefont {Swoboda}, \citenamefont {Vogel}, \citenamefont
  {Matsuyama},\ and\ \citenamefont {Meier}}]{kuhlmann_2013}%
  \BibitemOpen
  \bibfield  {author} {\bibinfo {author} {\bibfnamefont {N.}~\bibnamefont
  {Kuhlmann}}, \bibinfo {author} {\bibfnamefont {C.}~\bibnamefont {Swoboda}},
  \bibinfo {author} {\bibfnamefont {A.}~\bibnamefont {Vogel}}, \bibinfo
  {author} {\bibfnamefont {T.}~\bibnamefont {Matsuyama}}, \ and\ \bibinfo
  {author} {\bibfnamefont {G.}~\bibnamefont {Meier}},\ }\href@noop {}
  {\bibfield  {journal} {\bibinfo  {journal} {Phys. Rev. B}\ }\textbf {\bibinfo
  {volume} {87}},\ \bibinfo {pages} {104409} (\bibinfo {year}
  {2013})}\BibitemShut {NoStop}%
\bibitem [{\citenamefont {Lehndorff}\ \emph {et~al.}(2008)\citenamefont
  {Lehndorff}, \citenamefont {B{\"u}rgler}, \citenamefont {Kakay},
  \citenamefont {Hertel},\ and\ \citenamefont {Schneider}}]{lehndorff_2008}%
  \BibitemOpen
  \bibfield  {author} {\bibinfo {author} {\bibfnamefont {R.}~\bibnamefont
  {Lehndorff}}, \bibinfo {author} {\bibfnamefont {D.~E.}\ \bibnamefont
  {B{\"u}rgler}}, \bibinfo {author} {\bibfnamefont {A.}~\bibnamefont {Kakay}},
  \bibinfo {author} {\bibfnamefont {R.}~\bibnamefont {Hertel}}, \ and\ \bibinfo
  {author} {\bibfnamefont {C.~M.}\ \bibnamefont {Schneider}},\ }\href@noop {}
  {\bibfield  {journal} {\bibinfo  {journal} {IEEE Trans. Magn.}\ }\textbf
  {\bibinfo {volume} {44}},\ \bibinfo {pages} {1951} (\bibinfo {year}
  {2008})}\BibitemShut {NoStop}%
\bibitem [{\citenamefont {Krause}\ \emph {et~al.}(2007)\citenamefont {Krause},
  \citenamefont {Berbil-Bautista}, \citenamefont {Herzog}, \citenamefont
  {Bode},\ and\ \citenamefont {Wiesendanger}}]{krause_2007}%
  \BibitemOpen
  \bibfield  {author} {\bibinfo {author} {\bibfnamefont {S.}~\bibnamefont
  {Krause}}, \bibinfo {author} {\bibfnamefont {L.}~\bibnamefont
  {Berbil-Bautista}}, \bibinfo {author} {\bibfnamefont {G.}~\bibnamefont
  {Herzog}}, \bibinfo {author} {\bibfnamefont {M.}~\bibnamefont {Bode}}, \ and\
  \bibinfo {author} {\bibfnamefont {R.}~\bibnamefont {Wiesendanger}},\
  }\href@noop {} {\bibfield  {journal} {\bibinfo  {journal} {Science}\ }\textbf
  {\bibinfo {volume} {317}},\ \bibinfo {pages} {1537} (\bibinfo {year}
  {2007})}\BibitemShut {NoStop}%
\bibitem [{\citenamefont {Zabel}\ and\ \citenamefont {Farle}()}]{zabel_2013}%
  \BibitemOpen
  \bibfield  {author} {\bibinfo {author} {\bibfnamefont {H.}~\bibnamefont
  {Zabel}}\ and\ \bibinfo {author} {\bibfnamefont {M.}~\bibnamefont {Farle}},\
  }\href@noop {} {\bibinfo  {journal} {\emph{Magnetic Nanostructures: Spin
  Dynamics and Spin Transport} (Springer Berlin Heidelberg New York, 2013)}\
  }\BibitemShut {NoStop}%
\bibitem [{\citenamefont {Vedmedenko}\ and\ \citenamefont
  {Altwein}(2014)}]{vedmedenko_2014}%
  \BibitemOpen
\bibfield  {journal} {  }\bibfield  {author} {\bibinfo {author} {\bibfnamefont
  {E.~Y.}\ \bibnamefont {Vedmedenko}}\ and\ \bibinfo {author} {\bibfnamefont
  {D.}~\bibnamefont {Altwein}},\ }\href@noop {} {\bibfield  {journal} {\bibinfo
   {journal} {Phys. Rev. Lett.}\ }\textbf {\bibinfo {volume} {112}},\ \bibinfo
  {pages} {017206} (\bibinfo {year} {2014})}\BibitemShut {NoStop}%
\bibitem [{\citenamefont {Jensen}\ and\ \citenamefont
  {Mackintosh}()}]{jensen_1991}%
  \BibitemOpen
  \bibfield  {author} {\bibinfo {author} {\bibfnamefont {J.}~\bibnamefont
  {Jensen}}\ and\ \bibinfo {author} {\bibfnamefont {A.~R.}\ \bibnamefont
  {Mackintosh}},\ }\href@noop {} {\bibinfo  {journal} {\emph{Rare Earth
  Magnetism: Structures and Excitations} (Clarendon Press, Oxford, 1991)}\
  }\BibitemShut {NoStop}%
\bibitem [{\citenamefont {Blundell}()}]{blundell_2009}%
  \BibitemOpen
\bibfield  {journal} {  }\bibfield  {author} {\bibinfo {author} {\bibfnamefont
  {S.}~\bibnamefont {Blundell}},\ }\href@noop {} {\bibinfo  {journal}
  {\emph{Magnetism in Condensed Matter} (Oxford University Press, 2009)}\ ,\
  \bibinfo {pages} {p. 99}}\BibitemShut {NoStop}%
\bibitem [{\citenamefont {R{\"o}hlsberger}\ \emph {et~al.}(2002)\citenamefont
  {R{\"o}hlsberger}, \citenamefont {Thomas}, \citenamefont {Schlage},
  \citenamefont {Burkel}, \citenamefont {Leupold},\ and\ \citenamefont
  {R{\"u}ffer}}]{rohlsberger_2002}%
  \BibitemOpen
\bibfield  {journal} {  }\bibfield  {author} {\bibinfo {author} {\bibfnamefont
  {R.}~\bibnamefont {R{\"o}hlsberger}}, \bibinfo {author} {\bibfnamefont
  {H.}~\bibnamefont {Thomas}}, \bibinfo {author} {\bibfnamefont
  {K.}~\bibnamefont {Schlage}}, \bibinfo {author} {\bibfnamefont
  {E.}~\bibnamefont {Burkel}}, \bibinfo {author} {\bibfnamefont
  {O.}~\bibnamefont {Leupold}}, \ and\ \bibinfo {author} {\bibfnamefont
  {R.}~\bibnamefont {R{\"u}ffer}},\ }\href@noop {} {\bibfield  {journal}
  {\bibinfo  {journal} {Phys. Rev. Lett.}\ }\textbf {\bibinfo {volume} {89}},\
  \bibinfo {pages} {237201} (\bibinfo {year} {2002})}\BibitemShut {NoStop}%
\bibitem [{mag()}]{magnum}%
  \BibitemOpen
  \href@noop {} {\bibinfo  {journal}
  {http://micromagnum.informatik.uni-hamburg.de}\ }\BibitemShut {NoStop}%
\bibitem [{ani()}]{anisotropy}%
  \BibitemOpen
\bibfield  {journal} {  }\href@noop {} {\bibinfo  {journal} {As reported in
  Ref.~\onlinecite{seki_2013}, $K_\textnormal{1}$ is smaller compared to the
  experimental value
  \mbox{$K_\textnormal{1}=2.5\times10^{6}$~J/$\textnormal{m}^{3}$} because the
  one-dimensional model often leads to the underestimation of
  $K_\textnormal{1}$}\ }\BibitemShut {NoStop}%
\bibitem [{exp()}]{experiment}%
  \BibitemOpen
\bibfield  {journal} {  }\href@noop {} {\bibinfo  {journal} {In an experiment,
  one usually measures a coercive field $B_\textnormal{c}$ that can be found
  for a magnet with an uniaxial anisotropy via
  \mbox{$B_\textnormal{c}=2K/M_\textnormal{s}$}
  (Ref.~\onlinecite{getzlaff_2008}). In this study,
  \mbox{$B_\textnormal{c1}=800$~mT}, \mbox{$B_\textnormal{c2}=0$~mT} and
  \mbox{$B_\textnormal{c3}=74$~mT} are coercive fields for the hard FePt, soft
  Fe and hard Fe film, respectively. As reported in
  Ref.~\onlinecite{rohlsberger_2002}, \mbox{$B_\textnormal{c1}$} can be
  obtained up to 960~mT by heating the FePt film at a temperature $T=800$~K.
  The coercive field of the Fe film can be adjusted within a range of \mbox{1
  to 100~mT} via the direction of the sputtered Fe atoms relative to the
  orientation of the substrate surface using oblique incidence deposition (OID)
  at room temperature (see Ref.~\onlinecite{schlage_2014})}\ }\BibitemShut
  {NoStop}%
\bibitem [{\citenamefont {Fullerton}\ \emph {et~al.}(1997)\citenamefont
  {Fullerton}, \citenamefont {Jiang}, \citenamefont {Grimsditch}, \citenamefont
  {Sowers},\ and\ \citenamefont {Bader}}]{fullerton_1997}%
  \BibitemOpen
\bibfield  {journal} {  }\bibfield  {author} {\bibinfo {author} {\bibfnamefont
  {E.~E.}\ \bibnamefont {Fullerton}}, \bibinfo {author} {\bibfnamefont {J.~S.}\
  \bibnamefont {Jiang}}, \bibinfo {author} {\bibfnamefont {M.}~\bibnamefont
  {Grimsditch}}, \bibinfo {author} {\bibfnamefont {C.~H.}\ \bibnamefont
  {Sowers}}, \ and\ \bibinfo {author} {\bibfnamefont {S.~D.}\ \bibnamefont
  {Bader}},\ }\href@noop {} {\bibfield  {journal} {\bibinfo  {journal} {Phys.
  Rev. B}\ }\textbf {\bibinfo {volume} {58}},\ \bibinfo {pages} {12193}
  (\bibinfo {year} {1997})}\BibitemShut {NoStop}%
\bibitem [{\citenamefont {Stancil}\ and\ \citenamefont
  {Prabhakar}()}]{stancil_2009}%
  \BibitemOpen
  \bibfield  {author} {\bibinfo {author} {\bibfnamefont {D.~D.}\ \bibnamefont
  {Stancil}}\ and\ \bibinfo {author} {\bibfnamefont {A.}~\bibnamefont
  {Prabhakar}},\ }\href@noop {} {\bibinfo  {journal} {\emph{Spin Waves: Theory
  and Applications} (Springer Science+Business Media, LCC, 2009)}\
  }\BibitemShut {NoStop}%
\bibitem [{\citenamefont {Seki}\ \emph {et~al.}(2013)\citenamefont {Seki},
  \citenamefont {Utsumiya}, \citenamefont {Nozaki}, \citenamefont {Imamura},\
  and\ \citenamefont {Takanashi}}]{seki_2013}%
  \BibitemOpen
\bibfield  {journal} {  }\bibfield  {author} {\bibinfo {author} {\bibfnamefont
  {T.}~\bibnamefont {Seki}}, \bibinfo {author} {\bibfnamefont {K.}~\bibnamefont
  {Utsumiya}}, \bibinfo {author} {\bibfnamefont {Y.}~\bibnamefont {Nozaki}},
  \bibinfo {author} {\bibfnamefont {H.}~\bibnamefont {Imamura}}, \ and\
  \bibinfo {author} {\bibfnamefont {K.}~\bibnamefont {Takanashi}},\ }\href@noop
  {} {\bibfield  {journal} {\bibinfo  {journal} {Nature Commun.}\ }\textbf
  {\bibinfo {volume} {4}},\ \bibinfo {pages} {1726} (\bibinfo {year}
  {2013})}\BibitemShut {NoStop}%
\bibitem [{\citenamefont {Swoboda}\ \emph {et~al.}(2013)\citenamefont
  {Swoboda}, \citenamefont {Kuhlmann}, \citenamefont {Martens}, \citenamefont
  {Vogel},\ and\ \citenamefont {Meier}}]{swoboda_2013}%
  \BibitemOpen
  \bibfield  {author} {\bibinfo {author} {\bibfnamefont {C.}~\bibnamefont
  {Swoboda}}, \bibinfo {author} {\bibfnamefont {N.}~\bibnamefont {Kuhlmann}},
  \bibinfo {author} {\bibfnamefont {M.}~\bibnamefont {Martens}}, \bibinfo
  {author} {\bibfnamefont {A.}~\bibnamefont {Vogel}}, \ and\ \bibinfo {author}
  {\bibfnamefont {G.}~\bibnamefont {Meier}},\ }\href@noop {} {\bibfield
  {journal} {\bibinfo  {journal} {J. Appl. Phys.}\ }\textbf {\bibinfo {volume}
  {114}},\ \bibinfo {pages} {043905} (\bibinfo {year} {2013})}\BibitemShut
  {NoStop}%
\bibitem [{\citenamefont {Schlage}\ and\ \citenamefont
  {R{\"o}hlsberger}(2013)}]{schlage_2013}%
  \BibitemOpen
  \bibfield  {author} {\bibinfo {author} {\bibfnamefont {K.}~\bibnamefont
  {Schlage}}\ and\ \bibinfo {author} {\bibfnamefont {R.}~\bibnamefont
  {R{\"o}hlsberger}},\ }\href@noop {} {\bibfield  {journal} {\bibinfo
  {journal} {J. Elect. Spec. Rel. Phen.}\ }\textbf {\bibinfo {volume} {189}},\
  \bibinfo {pages} {187} (\bibinfo {year} {2013})}\BibitemShut {NoStop}%
\bibitem [{\citenamefont {Getzlaff}()}]{getzlaff_2008}%
  \BibitemOpen
  \bibfield  {author} {\bibinfo {author} {\bibfnamefont {M.}~\bibnamefont
  {Getzlaff}},\ }\href@noop {} {\bibinfo  {journal} {\emph{Fundamentals of
  Magnetism}, (Springer Berlin Heidelberg New York, 2008)}\ ,\ \bibinfo {pages}
  {p. 191}}\BibitemShut {NoStop}%
\bibitem [{\citenamefont {Schlage}\ \emph {et~al.}()\citenamefont {Schlage},
  \citenamefont {Erb}, \citenamefont {R{\"o}hlsberger}, \citenamefont {Wille},
  \citenamefont {Schumacher},\ and\ \citenamefont {Bocklage}}]{schlage_2014}%
  \BibitemOpen
\bibfield  {journal} {  }\bibfield  {author} {\bibinfo {author} {\bibfnamefont
  {K.}~\bibnamefont {Schlage}}, \bibinfo {author} {\bibfnamefont
  {D.}~\bibnamefont {Erb}}, \bibinfo {author} {\bibfnamefont {R.}~\bibnamefont
  {R{\"o}hlsberger}}, \bibinfo {author} {\bibfnamefont {H.-C.}\ \bibnamefont
  {Wille}}, \bibinfo {author} {\bibfnamefont {D.}~\bibnamefont {Schumacher}}, \
  and\ \bibinfo {author} {\bibfnamefont {L.}~\bibnamefont {Bocklage}},\
  }\href@noop {} {\bibinfo  {journal} {\emph{Method of producing a multilayer
  magnetoelectronic device}, European Patent, P91642 (2014)}\ }\BibitemShut
  {NoStop}%
\end{thebibliography}%

\end{document}